# On the scalability and message count of Trickle-based broadcasting schemes

Thomas M. M. Meyfroyt[1] · Sem C. Borst[1] ·
Onno J. Boxma[1] · Dee Denteneer[2]

**Abstract** As the use of wireless sensor networks increases, the need for efficient and reliable broadcasting algorithms grows. Ideally, a broadcasting algorithm should have the ability to quickly disseminate data, while keeping the number of transmissions low. In this paper, we analyze the popular Trickle algorithm, which has been proposed as a suitable communication protocol for code maintenance and propagation in wireless sensor networks. We show that the broadcasting process of a network using Trickle can be modeled by a Markov chain and that this chain falls under a class of Markov chains, closely related to residual lifetime distributions. It is then shown that this class of Markov chains admits a stationary distribution of a special form. These results are used to analyze the Trickle algorithm and its message count. Our results prove conjectures made in the literature concerning the effect of a listen-only period. Besides providing a mathematical analysis of the algorithm, we propose a generalized version of Trickle, with an additional parameter defining the length of a listen-only period.

✉ Thomas M. M. Meyfroyt
t.m.m.meyfroyt@tue.nl

Sem C. Borst
s.c.borst@tue.nl

Onno J. Boxma
o.j.boxma@tue.nl

Dee Denteneer
dee.denteneer@philips.com

[1] Eindhoven University of Technology, P.O. Box 513, 5600 MB Eindhoven, The Netherlands

[2] Philips Research, HTC 34, 5656 AE Eindhoven, The Netherlands

**Mathematics Subject Classification**　60J05 · 60J20 · 90B18

## 1 Introduction

Wireless sensor networks (WSNs) have become more and more popular in the last few years and have many applications [1]. These networks consist of compact, inexpensive sensor units that can communicate with each other by wireless transmissions. They require efficient and reliable communication protocols that can quickly propagate new information, while keeping the number of transmissions low, in order to conserve energy and maximize the lifetime of the network. Several data dissemination protocols have been proposed in recent years for this purpose [3,6,13,25].

In [13], the Trickle algorithm has been proposed in order to effectively and efficiently distribute and maintain code in wireless sensor networks. Trickle relies on a "polite gossip" policy to quickly propagate updates across a network of nodes. It uses a counter method to reduce the number of redundant transmissions in a network and to prevent a broadcast storm [17]. This makes Trickle also a very energy-efficient and popular method of maintaining a sensor network. The algorithm has been standardized by the IETF as the mechanism that regulates the transmission of the control messages used to create the network graph in the IPv6 Routing Protocol for Low power and Lossy Networks (RPL) [24]. Additionally, it is used in the Multicast Protocol for Low power and Lossy Networks (MPL), which provides IPv6 multicast forwarding in constrained networks and is currently being standardized [10]. Since the algorithm has such a broad applicability, the definition of Trickle has been documented in its own IETF RFC 6206 [12].

Because of the popularity of the algorithm, it is crucial to gain insight into how the various Trickle parameters influence network performance measures, such as energy consumption, available bandwidth, and latency. It is clear that such insights are crucial for optimizing the performance of wireless sensor networks. However, there are only a few results in that regard, most of them obtained via simulation studies [4,13,14].

Some analytical results are obtained in [2,11,14,15]. First, in [14], qualitative results are provided for the scalability of the Trickle algorithm, but a complete analysis is not given. More specifically, the authors of [14] state that in a lossless single-cell network without a listen-only period, the expected number of messages per time interval scales as $\mathcal{O}(\sqrt{n})$. Here $n$ is the number of nodes in the network and, adopting the terminology of [14], single-cell means that all the nodes in the network are within communication range from each other. Supposedly, introducing a listen-only period bounds the expected number of transmissions by a constant. The analysis in our paper confirms these claims and provides more explicit results. Secondly, in [11], a model is developed for estimating the message count of the Trickle algorithm in a multi-cell network, assuming a uniformly random spatial distribution of the nodes in the network. However, the influence of specific Trickle parameters on the message count is not explicitly addressed. Lastly, in [2,15], analytical models are developed for the time it takes the Trickle algorithm to update a network. To the best of the authors' knowledge, no other ana-lytical models for the performance of the Trickle algorithm have been published yet.

The goal of this paper is to develop and analyze stochastic models describing the Trickle algorithm and gain insight in how the Trickle parameters influence inter-transmission times and the message count. These insights could help optimize the energy-efficiency of the algorithm and consequently the lifetime of wireless sensor networks. Furthermore, knowing how the inter-transmission times depend on the various parameters could help prevent hidden-node problems in a network and optimize the capacity of wireless sensor networks. Additionally, our models are relevant for the analysis of communication protocols that build upon Trickle, such as RPL [24], MPL [10], CTP [8] and Deluge [9], and could give insight into their performance.

As key contributions of this paper, we first propose a generalized version of the Trickle algorithm by introducing a new parameter $\eta$, defining the length of a listen-only period. This addition proves to be useful for optimizing the design and usage of the algorithm. Furthermore, we derive the distribution of an inter-transmission time and the joint distribution of consecutive inter-transmission times for large-scale single-cell networks. We show how they depend on the Trickle parameters and investigate their asymptotic behavior. Additionally, we show that in a single-cell network without a listen-only period, the expected number of transmissions per time interval is unbounded and grows as $\sqrt{2n}\Gamma\left[\frac{k+1}{2}\right]/\Gamma\left[\frac{k}{2}\right]$, where $k$ is called the redundancy constant and $n$ the number of nodes as before. When a listen-only period of $\eta > 0$ is introduced, the message count is bounded by $k/\eta$ from above. We then use the results from the single-cell analysis to develop an approximation for the transmission count in multi-cell networks. All our results are compared and validated with simulation results and prove to be accurate, even for networks consisting of relatively few nodes.

Lastly, as an additional contribution of a more theoretical nature, we analyze a general class of Markov chains, closely related to residual lifetime distributions and provide the general stationary distribution for this class of chains. Compared with [16], which the present paper is based on, this contribution is the most important addition.

### 1.1 Organization of the paper

The remainder of this paper is organized as follows. In Sect. 2, we give a detailed description of the Trickle algorithm. Furthermore, we introduce a new parameter defining the length of a listen-only period and discuss its relevance. Then, in Sect. 3, we first briefly list the main results, before presenting the details of the model and its analysis. In Sect. 4, we develop a mathematical model describing the behavior of the Trickle algorithm in large-scale single-cell networks. This is followed by an analysis of a special class of Markov chains in Sect. 5 and we use the results from this analysis to analyze our original model in Sect. 6. In Sect. 7, we then validate our findings with simulations. In Sect. 8, we use the results from Sect. 6 to approximate the message count in multi-cell networks and we again compare our approximations with simulation results. In Sect. 9, we make some concluding remarks.

## 2 The Trickle algorithm

The Trickle algorithm has two main goals. First, whenever a new update enters the network, it must be propagated quickly throughout the network. Secondly, when there is no new update in the network, communication overhead has to be kept to a minimum.

The Trickle algorithm achieves this by using a "polite gossip" policy. Nodes divide time into intervals of varying length. During each interval, a node will broadcast its current information if it has heard fewer than, say, $k$ other nodes transmit the same information during that interval, in order to check if its information is up to date. If it has recently heard at least $k$ other nodes transmit the same information it currently has, it will stay quiet, assuming there is no new information to be received. Additionally, it will increase the length of its intervals, decreasing its broadcasting rate. Whenever a node receives an update or hears outdated information, it will reduce the length of its intervals, increasing its broadcasting rate, in order to quickly update nodes that have outdated information. This way inconsistencies are detected and resolved quickly, while keeping the number of transmissions low.

### 2.1 Algorithm description

We now describe the Trickle algorithm in its most general form (see also [13]). The algorithm has four parameters:

- A threshold value $k$, called the redundancy constant.
- The maximum interval length $\tau_h$.
- The minimum interval length $\tau_l$.
- The listen-only parameter $\eta$, defining the length of a listen-only period.

Furthermore, each node in the network has its own timer and keeps track of three variables:

- The current interval length $\tau$.
- A counter $c$, counting the number of messages heard during an interval.
- A broadcasting time $\theta$ during the current interval.

The behavior of each node is described by the following set of rules:

1. At the start of a new interval, a node resets its timer and counter $c$ and sets $\theta$ to a value in $[\eta\tau, \tau]$ uniformly at random.
2. When a node hears a message that is consistent with the information it has, it increments $c$ by 1.
3. When a node's timer hits time $\theta$, the node broadcasts its message if $c < k$.
4. When a node's timer hits time $\tau$, it doubles its interval length $\tau$ up to $\tau_h$ and starts a new interval.
5. When a node hears a message that is inconsistent with its own information, then if $\tau > \tau_l$ it sets $\tau$ to $\tau_l$ and starts a new interval, otherwise it does nothing.

In Fig. 1, an example is depicted of a network consisting of three nodes using the Trickle algorithm with $k = 1$ and $\tau = \tau_h$ for all nodes. During the first interval, node 3 is the first node that attempts to broadcast and consequently it is successful. The broadcasts of nodes 1 and 2 during that interval are then suppressed. During the second interval, the broadcast of node 2 suppresses the other broadcasts.

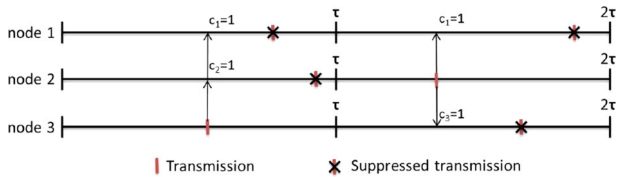

**Fig. 1** Example of three synchronized nodes using the Trickle algorithm

Note that in the example in Fig. 1, the intervals of the three nodes are synchronized. However, in general, the times at which nodes start their intervals need not be synchronized. In a synchronized network, all the nodes start their intervals at the same time, while in an unsynchronized network, this is not necessarily the case. In practice, networks will generally not be synchronized, since synchronization requires additional communication and consequently imposes energy overhead. Furthermore, as nodes get updated and start new intervals, they automatically lose synchronicity.

### 2.2 The listen-only parameter $\eta$

Note that the parameter $\eta$ is not introduced in the description of the Trickle algorithm in [12] and [13]. We have added this parameter ourselves, so we can analyze a more general version of the Trickle algorithm. The authors of [13] propose always using a listen-only period of half an interval, i.e., $\eta = \frac{1}{2}$, because of the so-called short-listen problem, which is discussed in the same paper. When no listen-only period is used, i.e., $\eta = 0$, sometimes nodes will broadcast soon after the beginning of their interval, listening for only a short time, before anyone else has a chance to speak up. If we have a perfectly synchronized network, this does not give a problem, because the first $k$ transmissions will simply suppress all the other broadcasts during that interval. However, in an unsynchronized network, if a node has a short listening period, it might broadcast just before another node starts its interval and that node possibly also has a short listening period. This possibly leads to a lot of redundant messages and is referred to as the short-listen problem.

In [13], it is claimed that not having a listen-only period and $k = 1$ makes the number of messages per time interval scale as $\mathcal{O}(\sqrt{n})$, due to the short-listen problem. When a listen-only period of $\tau/2$ is used, the expected number of messages per interval is supposedly bounded by 2, resolving the short-listen problem and improving scalability.

However, introducing a listen-only period also has its disadvantages. Firstly, when a listen-only period of $\tau/2$ is used, newly updated nodes will always have to wait for a period of at least $\tau_l/2$ before attempting to propagate the received update. Consequently, in an $m$-hop network, the end-to-end delay is at least $m\frac{\tau_l}{2}$. Hence, a listen-only period greatly affects the speed at which the Trickle algorithm can propagate updates. To prevent this, it has been proposed in [15] to dynamically adjust $\eta$, based on how new the current information of a node is, greatly increasing propagation speed.

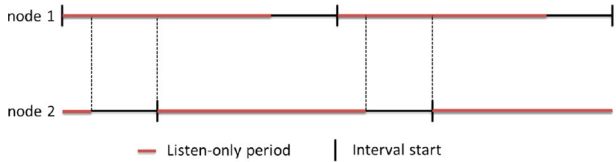

**Fig. 2** One node carries the complete transmission load of the network

Secondly, introducing a listen-only period has a negative effect on the load distribution. This is illustrated by Fig. 2, where node 2's broadcasting intervals completely overlap with node 1's listen-only periods and vice versa. Consequently, one node will always transmit and suppress the other node's transmissions, depending on which node first starts broadcasting. The probability of having an uneven load distribution increases as $\eta$ increases.

For these reasons, one might want to consider using a listen-only period of different lengths, which raises the question of what length is optimal. Therefore, we have added the parameter $\eta$, which allows us to investigate the effect of using a listen-only period of general length on the message count.

## 3 Main results for single-cell networks

In this section, we first briefly list the main results concerning the message count and scalability of the Trickle algorithm in single-cell networks. In the following sections, we will then discuss the analytical model and its analysis leading to these results.

We consider a steady-state regime, where all nodes are up to date and their interval lengths have settled to $\tau = \tau_h$ and without loss of generality, we will assume that $\tau_h = 1$. In this steady-state regime, as dictated by the Trickle algorithm, nodes still need to repeat the information they have, in order to be able to detect possible inconsistencies within the network. However, if no new information enters the network for a long period of time, all communication during this time is essentially redundant; hence, message count is the critical performance measure. When new information enters the network, nodes should quickly disseminate it and return to this steady-state regime.

We denote by $N^{(k,n)}$ the number of transmissions during an interval of length $\tau_h$ for a given threshold value $k$ in a single-cell network consisting of $n$ nodes. Additionally, let us denote an inter-transmission time for a given value of $k$ and cell-size $n$ by $T^{(k,n)}$. Whenever we write $a(n) \sim b(n) + c$, where $c$ is a constant, this means $a(n) - c$ is approximately equal to $b(n)$ as $n$ grows large. More formally

$$a(n) \sim b(n) + c \text{ denotes } \lim_{n \to \infty} (a(n) - c)/b(n) = 1.$$

First we look at the case $k = 1$. We show that the cumulative distribution function of $T^{(1,n)}$ for large $n$ behaves as

$$F^{(1,n)}(t) = \begin{cases} 0, & t < \eta, \\ 1 - e^{-\frac{n}{2}\frac{(t-\eta)^2}{1-\eta}}, & t \geq \eta. \end{cases} \tag{1}$$

This lets us deduce that

$$\mathbb{E}\left[T^{(1,n)}\right] \sim \eta + \sqrt{\frac{\pi(1-\eta)}{2n}}. \qquad (2)$$

We conclude that $\mathbb{E}[N^{(1,n)}] \sim \sqrt{\frac{2n}{\pi}}$, when $\eta = 0$. This proves the claim from [13] that when no listen-only period is used, $\mathbb{E}[N^{(1,n)}] = \mathcal{O}(\sqrt{n})$, and shows that the pre-factor is $\sqrt{\frac{2}{\pi}}$. Furthermore, when $\eta > 0$, Trickle scales well and $\mathbb{E}[N^{(1,n)}] \uparrow \frac{1}{\eta}$, with a convergence rate of $\sqrt{n}$, proving another claim from [13].

For the case $k \geq 2$, we first derive the density function for the distribution of $k-1$ consecutive inter-broadcasting times. We then use this result to deduce that the density function of the distribution of $T^{(k,n)}$ for large $n$ behaves as

$$f^{(k,n)}(t) = \frac{C_{(k,n)}}{(k-2)!} \int_0^\infty \lambda(t \mid v) v^{k-2} \exp\left[-\frac{n(t+v-\eta)^2}{2(1-\eta)}\right] dv. \qquad (3)$$

Here

$$\lambda(t \mid v) = \begin{cases} 0, & t+v < \eta, \\ \frac{n}{1-\eta}(t+v-\eta), & t+v \geq \eta, \end{cases} \qquad (4)$$

and

$$C_{(k,n)} = \left(\frac{\eta^{k-1}}{(k-1)!} + \int_\eta^\infty \frac{t^{k-2}}{(k-2)!} \exp\left[-\frac{n(t-\eta)^2}{2(1-\eta)}\right] dt\right)^{-1}. \qquad (5)$$

Additionally, we find for the $j$th moment of $T^{(k,n)}$:

$$\mathbb{E}\left[\left(T^{(k,n)}\right)^j\right] \sim j! \frac{C_{(k,n)}}{C_{(k+j,n)}}. \qquad (6)$$

Hence, for $\eta = 0$, $\mathbb{E}[N^{(k,n)}] \sim \sqrt{2n}\,\Gamma\left[\frac{k+1}{2}\right]/\Gamma\left[\frac{k}{2}\right]$, which is again $\mathcal{O}(\sqrt{n})$. Moreover, when $\eta > 0$, $\mathbb{E}[N^{(k,n)}] \uparrow \frac{k}{\eta}$, as $n \to \infty$, with a convergence rate of $\sqrt{n}$.

We then use these results to derive the asymptotic distributions of inter-broadcasting times. When $\eta > 0$ and $k \geq 2$, we show that

$$\frac{1}{\eta} T^{(k,n)} \xrightarrow{d} \text{Beta}(1, k-1), \text{ as } n \to \infty, \qquad (7)$$

and

$$\frac{k}{\eta} T^{(k,n)} \xrightarrow{d} \text{Exp}(1), \text{ as } n \to \infty \text{ and } k \to \infty. \qquad (8)$$

For the case $\eta = 0$, we show that the density function $f^{(k)}(t)$ of the limiting distribution of $\sqrt{\frac{n}{2}}T^{(k,n)}$ as $n \to \infty$ satisfies

$$f^{(k)}(t) = \frac{k-2}{k-3} f^{(k-2)}(t) - \frac{\Gamma\left[\frac{k}{2}\right]}{\Gamma\left[\frac{k-1}{2}\right]} \frac{2t}{k-3} f^{(k-1)}(t), \qquad (9)$$

where

$$f^{(2)}(t) = \frac{2}{\sqrt{\pi}} e^{-t^2},$$
$$f^{(3)}(t) = \sqrt{\pi}\,\mathrm{erfc}(t).$$

Lastly we show that

$$\sqrt{nk}\,T^{(k,n)} \xrightarrow{d} \mathrm{Exp}(1), \text{ as } n \to \infty \text{ and } k \to \infty. \qquad (10)$$

## 4 Modeling the broadcasting process

In this section, we develop a mathematical model describing the Trickle broadcasting process, which allows us to analyze the message count in single-cell networks. Suppose we have a single-cell network consisting of $n$ nodes which are all perfect receivers and transmitters. Furthermore, we assume that all nodes are up to date and $\tau = \tau_h = 1$ for all nodes. Lastly, we assume that the interval skew of the nodes is uniformly distributed, meaning each node has one interval starting at some time in the interval $[0, 1)$ uniformly at random.

In order to analyze the message count, we treat the process of nodes attempting to broadcast as a Poisson process with rate $n$. This assumption is motivated by the following lemma, which shows that the properly scaled process of nodes attempting to broadcast behaves as a Poisson process with rate 1 as $n$ grows large.

**Lemma 1** *Let $N_n$ be the point process of times that nodes attempt to broadcast in a single cell consisting of $n$ nodes with $\eta \in [0, 1]$. Then if we dilate the timescale by a factor $n$, the process $N_n$ converges weakly to a Poisson process with rate 1 as $n$ grows large.*

*Proof* See Appendix. □

Let us denote the PDF and CDF of an inter-transmission time $T_P^{(k,n)}$ for this process with Poisson broadcasting attempts by $f^{(k,n)}(t)$ and $F^{(k,n)}(t)$, respectively. Recall that by $T^{(k,n)}$ we denote an inter-transmission time for a given value of $k$ and cell-size $n$ for the original Trickle broadcasting process, which is not Poisson. We will use the fact that the distribution function of $T_P^{(k,n)}$ will tend to provide an accurate approximation for that of $T^{(k,n)}$ for large $n$ (because of Lemma 1), but the derivation of explicit error bounds or limit theorems is outside the scope of the present paper. It is important to keep in mind that we are analyzing the Poisson broadcasting process, and all results for $T^{(k,n)}$ (and $N^{(k,n)}$) are only true asymptotically, for $n \to \infty$.

### 4.1 $k = 1$

We first consider the case $k = 1$. Suppose at time 0, a broadcast occurs. Now assume at time $t$, a node ends its listening period and attempts to broadcast. In order for the node's transmission not to be suppressed, it must have started its listening period after time 0. If it started listening before this time, it would have heard the transmission at time 0 and its broadcast would have been suppressed. This means that the node's broadcast will be successful if its corresponding timer was smaller than $t$. Since broadcasting times are picked uniformly in $[\eta, 1]$, this probability is 0 if $t < \eta$ and $\frac{t-\eta}{1-\eta}$ otherwise. Hence, if we define $\lambda(t)$ to be the instantaneous rate at which successful broadcasts occur, we can write

$$\lambda(t) = \begin{cases} 0, & t < \eta, \\ \frac{n}{1-\eta}(t - \eta), & t \geq \eta. \end{cases} \quad (11)$$

It is well known that the hazard rate $\lambda(t) = \frac{f^{(1,n)}(t)}{1 - F^{(1,n)}(t)}$ uniquely determines $F^{(1,n)}(t) = \mathbb{P}[T_P^{(1,n)} \leq t]$ (see [7], Theorem 2.1):

$$F^{(1,n)}(t) = 1 - e^{-\int_0^t \lambda(u)\, du} = \begin{cases} 0, & t < \eta, \\ 1 - e^{-\frac{n}{2}\frac{(t-\eta)^2}{1-\eta}}, & t \geq \eta. \end{cases} \quad (12)$$

Hence, $\sqrt{\frac{n}{1-\eta}}(T_P^{(1,n)} - \eta)$ is a Rayleigh distributed random variable with scale parameter $\sigma = 1$. Therefore,

$$\mathbb{E}\left[T^{(1,n)}\right] \sim \mathbb{E}\left[T_P^{(1,n)}\right] = \eta + \sqrt{\frac{\pi(1-\eta)}{2n}}. \quad (13)$$

We conclude

$$\mathbb{E}\left[N^{(1,n)}\right] = \left(\mathbb{E}\left[T^{(1,n)}\right]\right)^{-1} \sim \left(\sqrt{\frac{\pi(1-\eta)}{2n}} + \eta\right)^{-1}. \quad (14)$$

Hence, for the case $\eta = 0$, we find $\mathbb{E}[N^{(1,n)}] \sim \sqrt{\frac{2n}{\pi}}$. This proves the claim from [13] that $\mathbb{E}[N^{(1,n)}] = \mathcal{O}(\sqrt{n})$ when no listen-only period is used. For $\eta > 0$, we get from (14) that

$$\mathbb{E}\left[N^{(1,n)}\right] \sim \frac{1}{\eta} - \frac{1}{\eta^2}\sqrt{\frac{\pi(1-\eta)}{2n}} + \mathcal{O}(n^{-1}). \quad (15)$$

This implies $\mathbb{E}[N^{(1,n)}] \uparrow \frac{1}{\eta}$ from below, proving the claim that introducing a listen-only period bounds the number of transmissions per interval by a constant.

### 4.2 $k \geq 2$

We now look at $k \geq 2$, starting by examining the case $k = 2$. We can apply similar reasoning as we did for the case $k = 1$. Suppose again that at time 0 a broadcast occurs and let $-T_{-1}$ be the time of the last broadcast before time 0. Now similarly to before, a broadcasting attempt at time $t$ will be successful if the corresponding node started its listening interval after time $-T_{-1}$. Hence, the instantaneous rate at which successful broadcasts occur conditioned on $T_{-1} = \nu$ is given by

$$\lambda(t \mid \nu) = \begin{cases} 0, & t + \nu < \eta, \\ \frac{n}{1-\eta}(t + \nu - \eta), & t + \nu \geq \eta. \end{cases} \quad (16)$$

Hence,

$$\mathbb{P}\left[T_P^{(2,n)} \leq t \mid T_{-1} = \nu\right] = 1 - \exp\left[-\int_0^t \lambda(u \mid \nu) du\right]$$

$$= \begin{cases} 0, & t + \nu < \eta, \\ 1 - \exp\left[-\frac{n(t+\nu-\eta)^2}{2(1-\eta)}\right], & \nu < \eta \wedge t + \nu \geq \eta, \\ 1 - \exp\left[-\frac{n(t^2/2 + t(\nu-\eta))}{1-\eta}\right], & \nu \geq \eta. \end{cases} \quad (17)$$

This tells us that for $k = 2$, the sequence of consecutive inter-transmission times $T_P^{(2,n)} = \{T_{P,i}^{(2,n)}\}_{i=0}^\infty$ forms a Markov chain with transition probabilities as in (17).

For the general case $k \geq 2$, the same reasoning applies. Suppose again that at time 0, a broadcast occurs and let $-T_{-(k-1)}$ be the time of the $(k-1)$th broadcast before time 0. Then a broadcasting attempt at time $t$ will be successful if the corresponding node started its listening interval after time $-T_{-(k-1)}$. Hence, the instantaneous rate at which broadcasts occur conditioned on $T_{-(k-1)} = \nu$ is again given by Eq. (16). Thus, in general, the sequence $T_P^{(k,n)} = \left\{\left(T_{P,i}^{(k,n)}, \ldots, T_{P,i+k-2}^{(k,n)}\right)\right\}_{i=0}^\infty$ forms a Markov chain. Moreover, as we will show in the next section, this Markov chain is a specific case of a class of Markov chains closely related to residual lifetime distributions.

## 5 Markov chains of residual lifetimes

In order to analyze the Markov chain $T_P^{(k,n)}$ and determine its steady-state distribution, it will help us to first study a more general class of Markov chains. The most important result of this section is the following theorem.

**Theorem 1** *Let $Y$ be a continuous random variable with support $(a, b)$ for some $a \geq 0$ and $b > 0$ and where we allow $b = \infty$. Denote by $F(y)$ the cumulative distribution function of $Y$. Let $X = \{X_i\}_{i=1}^\infty$ be an m-dependent sequence with probability transition function of the following specific form:*

$$\mathbb{P}[X_{n+1} \leq y \mid X_{n-m+1} = x_1, \ldots, X_n = x_m] = \mathbb{P}\left[Y \leq \sum_{j=1}^{m} x_j + y \,\bigg|\, Y \geq \sum_{j=1}^{m} x_j\right]. \tag{18}$$

Then, if $\mathbb{E}[Y^m] < \infty$,

$$\Pi_X(y) := \lim_{i \to \infty} \mathbb{P}[X_i \leq y] = 1 - \frac{m}{\mathbb{E}[Y^m]} \int_0^\infty (1 - F(s+y)) s^{m-1} ds. \tag{19}$$

*Remark 1* The distribution function in Eq. (19) can be interpreted as the stationary distribution of the $m$th iterated overshoot of a renewal process with inter-renewal distribution $F(\cdot)$: see [23]. That is, the residual of the residual of the residual... of $Y$ iterated $m$ times.

In preparation for the proof of Theorem 1, we shall first study the sequence $X_m = \{(X_i, \ldots, X_{i+m-1})\}_{i=1}^\infty$. Define $\bar{F}(y) = 1 - F(y)$ and let $\Lambda(y) = -\log[\bar{F}(y)]$ be the cumulative hazard function of $Y$. Note that we can write $F(y) = 1 - e^{-\Lambda(y)}$. Clearly, $X_m$ constitutes a Markov chain with state space $\mathcal{X} = \{x \in \mathbb{R}^m : \sum_{j=1}^m x_j \leq b \text{ and } x_i \geq 0 \text{ for } 1 \leq i \leq m\}$ and Markov transition function $P$ as in (18). That is

$$P((x_1, \ldots, x_m), (x_2, \ldots, x_m, dy)) = d\mathbb{P}\left[Y \leq \sum_{j=1}^{m} x_j + y \,\bigg|\, Y \geq \sum_{j=1}^{m} x_j\right]$$
$$= \left[d\Lambda\left(\sum_{j=1}^{m} x_j + y\right)\right] e^{-\Lambda\left(\sum_{j=1}^m x_j + y\right) + \Lambda\left(\sum_{j=1}^m x_j\right)}. \tag{20}$$

We show the following

**Theorem 2** *An invariant measure of the chain $X_m$ is given by*

$$\pi(x_1, \ldots, x_m) = C_m e^{-\Lambda\left(\sum_{j=1}^m x_j\right)} = C_m \bar{F}\left(\sum_{j=1}^{m} x_j\right), \tag{21}$$

*where $C_m$ is a constant.*

*Proof* Using (20) and (21) we find

$$\int_0^\infty \pi(x_1, \ldots, x_m) P((x_1, \ldots, x_m), (x_2, \ldots, x_m, dx_{m+1})) dx_1$$
$$= C_m \int_0^\infty \left[d\Lambda\left(\sum_{j=1}^{m+1} x_j\right)\right] e^{-\Lambda\left(\sum_{j=1}^{m+1} x_j\right)} dx_1$$

$$= C_m e^{-\Lambda\left(\sum_{j=2}^{m+1} x_j\right)} = C_m \bar{F}\left(\sum_{j=2}^{m+1} x_j\right) = \pi(x_2, \ldots, x_{m+1}).$$

□

If $\mathbb{E}[Y^m] < \infty$, then (21) can be normalized with a constant $C_m$ that satisfies

$$C_m = \left(\int_0^\infty \cdots \int_0^\infty e^{-\Lambda\left(\sum_{j=1}^{m} x_j\right)} \mathrm{d}x_m \cdots \mathrm{d}x_1\right)^{-1}$$
$$= \left(\int_0^\infty \frac{y^{m-1}}{(m-1)!} \bar{F}(y)\,\mathrm{d}y\right)^{-1} = \frac{m!}{\mathbb{E}[Y^m]}, \qquad (22)$$

where we have used the following lemma:

**Lemma 2** *Let $m \in \mathbb{N}$ and $G(x)$ be a positive real-valued integrable function. Assume that $\int_0^\infty x^m G(x)dx < \infty$, then*

$$\int_0^\infty \cdots \int_0^\infty G\left(\sum_{i=1}^{m+1} x_i\right) dx_1 \cdots dx_{m+1} = \int_0^\infty \frac{x^m}{m!} G(x) dx.$$

*Proof* By writing $x = \sum_{i=1}^{m+1} x_i$ and a change of variables, we can write

$$\int_0^\infty \cdots \int_0^\infty G\left(\sum_{i=1}^{m+1} x_i\right) dx_1 \cdots dx_{m+1}$$
$$= \int_0^\infty G(x) \int \cdots \int_{\substack{x_1+\cdots+x_m \leq x \\ x_i \geq 0 \text{ for } 1 \leq i \leq m}} dx_1 \cdots dx_m\, dx.$$

The inner $m$-tuple integral is equal to the volume of the $m$-dimensional simplex given by $\{(x_1, \cdots, x_m) \mid \sum_{i=1}^m x_i \leq x \text{ and } x_i \geq 0 \text{ for } 1 \leq i \leq m\}$, which is known to be $x^m/m!$ [19]. Applying this result completes the proof. □

*Remark 2* In general, the steady-state joint cumulative distribution function of $\mathbf{X}_m$ cannot be written as neatly as its density (21). However, one can show through induction that its multivariate Laplace transform is given by

$$\mathcal{L}_{\mathbf{X}_m}(s_1, \ldots, s_m) = C_m \sum_{i=1}^m \frac{1 - \mathcal{L}_Y(s_i)}{s_i} \frac{1}{\prod_{j \neq i}(s_j - s_i)}, \qquad (23)$$

where $\mathcal{L}_Y(s) = \int_0^\infty e^{-sy}\,\mathrm{d}F(y)$ is the Laplace-Stieltjes transform of $Y$.

In Theorem 2, we have established that the chain $X_m$ has an invariant measure $\pi$ as given in (21). We now proceed to show that the density function of $X_m$ also converges to $\pi$ as in (21) with normalization constant as given in (22) for any starting vector $x \in \mathcal{X}$, if $\mathbb{E}[Y^m] < \infty$.

Given the existence of a strictly positive finite invariant measure $\pi$, inspection of Theorem 4 in [18] shows that if the chain $X_m$ is aperiodic and $\phi$-irreducible, then the distribution of $X_m$ will converge to the distribution associated with the invariant measure given in (21).

It follows easily from the fact that $Y$ has support $(a, b)$ for some $a \geq 0$ and $b > 0$ that $X_m$ is aperiodic. Furthermore, $\phi$-irreducibility of the chain (i.e., having positive probability of reaching every set $A$ with $\phi(A) > 0$ from every state $x \in \mathcal{X}$, for some nonzero measure $\phi(\cdot)$) is also easy to verify. We can take $\phi(A) = \mu_L(A \cap \mathcal{X})$, where $\mu_L(\cdot)$ denotes the Lebesgue measure on $\mathbb{R}^m$. Starting from any $x \in \mathcal{X}$, the chain is able to move to the set $\{x \in \mathbb{R}^m : \min(x_1, \ldots, x_m) \geq a\}$ within $m$ steps, and from there it can reach any other set in $\mathcal{X}$ within $m$ steps. Hence, under the conditions of Theorem 1, $X_m$ is aperiodic and $\phi$-irreducible.

Now, with the stationary density function of $X_m$, we can also determine the associated steady-state density of the process $\Sigma = \left\{\sum_{j=0}^{m-1} X_{i+j}\right\}_{i=1}^{\infty}$, which we will denote by $\pi_\Sigma$. Using (21), we find

$$\pi_\Sigma(x) = \int_0^x \int_0^{x-x_1} \cdots \int_0^{x-\sum_{j=1}^{m-2} x_j} \pi\left(x_1, \ldots, x_{m-1}, x - \sum_{j=1}^{m-1} x_j\right) dx_{m-1} \cdots dx_1$$

$$= C_m \int_0^x \int_0^{x-x_1} \cdots \int_0^{x-\sum_{j=1}^{m-2} x_j} \bar{F}(x) \, dx_{m-1} \cdots dx_1 = \frac{m}{\mathbb{E}[Y^m]} x^{m-1} \bar{F}(x). \quad (24)$$

We are now in a position to prove Theorem 1.

*Proof of Theorem 1* We already derived that under the conditions of Theorem 1, the density of $X_m$ will converge to the normalized version of the expression given in Eq. (21). Consequently, the density of $\Sigma$ will converge to (24). Therefore, we can write the limiting stationary distribution of $X$ as

$$\Pi_X(y) = \lim_{i \to \infty} \mathbb{P}[X_i \leq y] = \int_0^\infty \pi_\Sigma(s) \mathbb{P}[Y \leq y + s \mid Y \geq s] ds$$

$$= \int_0^\infty \pi_\Sigma(s) \left(1 - e^{-\Lambda(s+y) + \Lambda(s)}\right) ds$$

$$= 1 - \frac{m}{\mathbb{E}[Y^m]} \int_0^\infty (1 - F(s+y)) s^{m-1} ds. \quad (25)$$

Hence, we have proven Theorem 1. □

Lastly, we use (19) to derive the moments of the stationary distribution of $X$. For all $j \in \mathbb{N}$ such that $\mathbb{E}[Y^{m+j}] < \infty$,

$$\int_0^\infty y^j \, d\Pi_X(y) = \frac{m}{\mathbb{E}[Y^m]} \int_0^\infty \int_0^\infty y^j s^{m-1} \, dF(s+y) \, dy$$

$$= \left[ \binom{m+j}{j} \right]^{-1} \frac{1}{\mathbb{E}[Y^m]} \int_0^\infty y^{m+j} \, dF(y)$$

$$= \left[ \binom{m+j}{j} \right]^{-1} \frac{\mathbb{E}[Y^{m+j}]}{\mathbb{E}[Y^m]}, \qquad (26)$$

where we have used the following lemma:

**Lemma 3** *Let $m \in \mathbb{N}$ and $G(x)$ be a positive real-valued integrable function. Assume that $\int_0^\infty x^{m+j+1} G(x) dx < \infty$, then*

$$\int_0^\infty \int_0^\infty x^j y^m G(x+y) \, dx \, dy = \frac{1}{m+1} \left[ \binom{m+j+1}{j} \right]^{-1} \int_0^\infty z^{m+j+1} G(z) \, dz.$$

*Proof* Write $z = x + y$, then

$$\int_0^\infty \int_0^\infty x^j y^m G(x+y) \, dx \, dy = \int_0^\infty \int_y^\infty (z-y)^j y^m G(z) \, dz \, dy$$

$$= \sum_{i=0}^j (-1)^i \binom{j}{i} \int_0^\infty \int_y^\infty z^{j-i} y^{m+i} G(z) \, dz \, dy$$

$$= \sum_{i=0}^j (-1)^i \binom{j}{i} \int_0^\infty z^{j-i} G(z) \int_0^z y^{m+i} \, dy \, dz$$

$$= \left( \sum_{i=0}^j (-1)^i \binom{j}{i} \frac{1}{1+i+m} \right) \int_0^\infty z^{1+m+j} G(z) \, dz$$

$$= \frac{1}{m+1} \left[ \binom{m+j+1}{j} \right]^{-1} \int_0^\infty z^{m+j+1} G(z) \, dz.$$

The last equality follows from a known identity involving the reciprocal of binomial coefficients (see [21], Corollary 2.2). □

We will now use the results from this section to analyze the Markov chain $T_P^{(k,n)}$ from Sect. 4.

## 6 Inter-transmission time distributions and message count

We return to the model from Section 4. We already deduced that the sequence $T_P^{(k,n)} = \left\{ \left( T_{P,i}^{(k,n)}, \ldots T_{P,i+k-2}^{(k,n)} \right) \right\}_{i=0}^\infty$ forms a Markov chain. Moreover, we showed that for $k \geq 2$,

$$\mathbb{P}\left[T_P^{(k,n)} \leq t \mid T_{-(k-1)} = v\right] = 1 - \exp\left[-\int_0^t \lambda(u \mid v) du\right], \quad (27)$$

with $\lambda(t \mid v)$ as in (16). Since $T_{-(k-1)}$ is actually the sum of the previous $k-1$ inter-broadcasting times, we find that (27) is of the form as in (18) with $m = k-1$. Further examination then gives that for this case $Y \sim T_P^{1,n}$ and $F(t)$ is given by (12). Now, since all the moments of $T_P^{1,n}$ are finite and $T_P^{1,n}$ has support $(\eta, \infty)$, Theorem 1 applies, and we can apply all the results from the previous Sect. to $T_P^{(k,n)}$.

Recall that by $f^{(k,n)}(t)$ and $F^{(k,n)}(t)$, we denote the steady-state PDF and CDF of a transmission time $T_P^{(k,n)}$, respectively. Additionally, let us denote the steady-state joint probability density function of $k-1$ consecutive inter-transmission times by $\tilde{f}^{(k,n)}(t_1, \ldots, t_{k-1})$ and let $f_\Sigma^{(k,n)}(s)$ be the steady-state probability density function of the sequence $\Sigma^{(k,n)} = \{\sum_{j=0}^{k-2} T_{P,i+j}^{(k,n)}\}_{i=0}^\infty$.

It then follows directly from (21) that

$$\tilde{f}^{(k,n)}(t_1,\ldots,t_{k-1}) = \begin{cases} C_{(k,n)}, & \sum_{i=1}^{k-1} t_i < \eta, \\ C_{(k,n)} \exp\left[-\frac{n}{2(1-\eta)}\left(\sum_{i=1}^{k-1} t_i - \eta\right)^2\right], & \sum_{i=1}^{k-1} t_i \geq \eta. \end{cases} \quad (28)$$

Here the normalization constant $C_{(k,n)}$ satisfies

$$C_{(k,n)} = \frac{(k-1)!}{\mathbb{E}\left[\left(T_P^{1,n}\right)^{k-1}\right]} = \left(\frac{\eta^{k-1}}{(k-1)!} + \int_\eta^\infty \frac{t^{k-2}}{(k-2)!} \exp\left[-\frac{n(t-\eta)^2}{2(1-\eta)}\right] dt\right)^{-1}. \quad (29)$$

Alternatively, by a change of variables and splitting the integral, we can write (29) in terms of a finite sum as

$$C_{(k,n)} = \left(\frac{\eta^{k-1}}{(k-1)!} + \frac{1}{2(k-2)!} \sum_{i=0}^{k-2} \binom{k-2}{i} \eta^{k-i-2} \left(\frac{2(1-\eta)}{n}\right)^{\frac{i+1}{2}} \Gamma\left[\frac{i+1}{2}\right]\right)^{-1},$$

which more clearly reveals the role of some of the parameters.

There are two important observations we can make regarding the constant $C_{(k,n)}$. First, for $\eta = 0$, (29) reduces to $C_{(k,n)} = (2n)^{\frac{k-1}{2}} \Gamma[k/2]/\sqrt{\pi}$. Secondly, for $\eta > 0$, we have that $C_{(k,n)} \downarrow \frac{(k-1)!}{\eta^{k-1}}$ as $n$ grows large.

Let us now look at the sequence $\Sigma^{(k,n)}$, i.e., the sequence of the sum of $k-1$ consecutive inter-transmissions times. Equation (24) immediately gives its steady-state probability density function:

$$f_\Sigma^{(k,n)}(s) = \begin{cases} \frac{C_{(k,n)}}{(k-2)!} s^{k-2}, & s < \eta, \\ \frac{C_{(k,n)}}{(k-2)!} s^{k-2} \exp\left[-\frac{n(s-\eta)^2}{2(1-\eta)}\right], & s \geq \eta. \end{cases} \quad (30)$$

Finally, Eq. (19) gives

$$F^{(k,n)}(t) = 1 - \frac{C_{(k,n)}}{(k-2)!} \int_0^\infty \left(1 - F^{(1,n)}(s+t)\right) s^{m-1} \, ds, \qquad (31)$$

and equivalently,

$$f^{(k,n)}(t) = \frac{C_{(k,n)}}{(k-2)!} \int_0^\infty f^{(1,n)}(s+t) s^{m-1} ds. \qquad (32)$$

Substitution of Eq. (16) reduces the right-hand side of (32) for $t < \eta$ to

$$\frac{n}{1-\eta} \frac{C_{(k,n)}}{(k-2)!} \int_{\eta-t}^\infty (t+v-\eta) v^{k-2} \exp\left[-\frac{n(t+v-\eta)^2}{2(1-\eta)}\right] dv, \qquad (33)$$

and for $t \geq \eta$, the right-hand side reduces to

$$\frac{n}{1-\eta} \frac{C_{(k,n)}}{(k-2)!} \int_0^\infty (t+v-\eta) v^{k-2} \exp\left[-\frac{n(t+v-\eta)^2}{2(1-\eta)}\right] dv. \qquad (34)$$

Lastly, we focus our attention on the moments of an inter-transmission time $T^{(k,n)}$. Applying (26), we find for the first moment of $T^{(k,n)}$

$$\mathbb{E}\left[T^{(k,n)}\right] \sim \mathbb{E}\left[T_P^{(k,n)}\right] = \frac{C_{(k,n)}}{C_{(k+1,n)}}. \qquad (35)$$

For the case $\eta = 0$, this implies

$$\mathbb{E}\left[N^{(k,n)}\right] = \left(\mathbb{E}\left[T^{(k,n)}\right]\right)^{-1} \sim \sqrt{2n} \frac{\Gamma\left[\frac{k+1}{2}\right]}{\Gamma\left[\frac{k}{2}\right]}, \qquad (36)$$

which is again $\mathcal{O}(\sqrt{n})$, as was conjectured in [13].

When a listen-only period is used, i.e., $\eta > 0$,

$$\mathbb{E}\left[N^{(k,n)}\right] = \left(\mathbb{E}\left[T^{(k,n)}\right]\right)^{-1} \sim \frac{k}{\eta} - \frac{k}{\eta^2} \sqrt{\frac{\pi(1-\eta)}{2n}} + \mathcal{O}(n^{-1}), \qquad (37)$$

implying that $\mathbb{E}[N^{(k,n)}] \uparrow \frac{k}{\eta}$ from below as $n \to \infty$. This implies that for $\eta > 0$, the expected number of messages per interval is bounded by a constant and shows that Trickle scales well.

Again using (26), we obtain the following expression for the $j$th moment of an inter-transmission time

$$\mathbb{E}\left[\left(T^{(k,n)}\right)^j\right] \sim \mathbb{E}\left[\left(T_P^{(k,n)}\right)^j\right] = j! \frac{C_{(k,n)}}{C_{(k+j,n)}}. \qquad (38)$$

For the special case $\eta = 0$, this reduces to

$$\mathbb{E}\left[\left(T^{(k,n)}\right)^j\right] \sim \frac{j!}{(2n)^{\frac{j}{2}}} \frac{\Gamma\left[\frac{k}{2}\right]}{\Gamma\left[\frac{k+j}{2}\right]}. \tag{39}$$

For the case $\eta > 0$, we deduce

$$\mathbb{E}\left[\left(T^{(k,n)}\right)^j\right] \to \frac{(k-1)!j!}{(k+j-1)!}\eta^j, \text{ as } n \to \infty. \tag{40}$$

### 6.1 Limiting distributions

The previous analysis allows us to determine the limiting distributions of $T^{(k,n)}$ as $n$ and $k$ grow large. We distinguish between two cases.

**Case 1:** $\eta > 0$.
First, Eq. (40) implies that for $\eta > 0$ and $k \geq 2$,

$$\frac{1}{\eta}T^{(k,n)} \xrightarrow{d} \text{Beta}(1, k-1), \text{ as } n \to \infty, \tag{41}$$

since all moments converge to the moments of the Beta distribution. For the density function of $T_P^{(k,n)}$ (and hence also that of $T^{(k,n)}$), this implies that as $n \to \infty$,

$$f^{(k,n)}(t) \to \begin{cases} \frac{(k-1)}{\eta}\left(1 - \frac{t}{\eta}\right)^{k-2}, & 0 \leq t \leq \eta, \\ 0, & \text{otherwise.} \end{cases} \tag{42}$$

Furthermore, since $k\text{Beta}(1, k) \xrightarrow{d} \text{Exp}(1)$ for $k \to \infty$,

$$\frac{k}{\eta}T^{(k,n)} \xrightarrow{d} \text{Exp}(1), \text{ as } n \to \infty \text{ and } k \to \infty. \tag{43}$$

We can relate this result to Equation (28). We know that the sum of $k$ consecutive inter-transmission times converges to $\eta$ as $n \to \infty$. Equation (28) then tells us that if we look at an interval of length $\eta$, $k - 1$ broadcasting times are distributed uniformly in this interval, when $n \to \infty$. Therefore, if we scale time by a factor $k$, we will see a Poisson process with intensity 1, as $k \to \infty$, and expect to see exponential inter-broadcasting times, which is in agreement with (43).

**Case 2**: $\eta = 0$.
For the case $\eta = 0$, we can write the density function $f^{(k)}(t)$ of the limiting distribution of $\sqrt{\frac{n}{2}}T^{(k,n)}$ for $k \geq 2$ after scaling Equation (34) as

$$f^{(k)}(t) = \frac{4}{\Gamma\left[\frac{k-1}{2}\right]} \int_0^\infty (t+v)v^{k-2} \exp\left[-(t+v)^2\right] dv$$

$$= \frac{4}{\Gamma\left[\frac{k-1}{2}\right]} t \int_0^\infty v^{k-2} \exp\left[-(t+v)^2\right] dv$$

$$- \frac{4}{\Gamma\left[\frac{k-1}{2}\right]} \int_0^\infty v^{k-1} \exp\left[-(t+v)^2\right] dv. \quad (44)$$

Alternatively, performing integration by parts gives

$$f^{(k)}(t) = \frac{2(k-1)}{\Gamma\left[\frac{k-1}{2}\right]} \int_0^\infty v^{k-3} \exp\left[-(t+v)^2\right] dv. \quad (45)$$

Combining (44) and (45), we find after some rewriting

$$f^{(k)}(t) = \frac{k-2}{k-3} f^{(k-2)}(t) - \frac{\Gamma\left[\frac{k}{2}\right]}{\Gamma\left[\frac{k-1}{2}\right]} \frac{2t}{k-3} f^{(k-1)}(t). \quad (46)$$

Direct computation gives

$$f^{(2)}(t) = \frac{2}{\sqrt{\pi}} e^{-t^2},$$

and

$$f^{(3)}(t) = \sqrt{\pi}\,\text{erfc}(t).$$

All other distribution functions then follow from Equation (46). Additionally, from Equation (39), we have

$$\mathbb{E}\left[\left(\sqrt{n}T^{(k,n)}\right)^j\right] \to \frac{j!}{2^{\frac{j}{2}}} \frac{\Gamma\left[\frac{k}{2}\right]}{\Gamma\left[\frac{k+j}{2}\right]}, \text{ as } n \to \infty. \quad (47)$$

Using the fact that

$$\lim_{k\to\infty} \frac{\Gamma\left[\frac{k}{2}\right]\left(\frac{k}{2}\right)^{\frac{j}{2}}}{\Gamma\left[\frac{k+j}{2}\right]} = 1,$$

we find

$$\mathbb{E}\left[\left(\sqrt{nk}T^{(k,n)}\right)^j\right] \to j!, \text{ as } n \to \infty \text{ and } k \to \infty. \quad (48)$$

This finally implies that

$$\sqrt{nk}\,T^{(k,n)} \xrightarrow{d} \text{Exp}(1), \text{ as } n \to \infty \text{ and } k \to \infty, \qquad (49)$$

since all moments converge to the moments of the exponential distribution.

The last result can be related to Lemma 1. If we look at the special case $k = n$, we know all broadcasting attempts will be successful. Hence, the process of nodes broadcasting will be the same as the process of nodes attempting to broadcast, from which we know it converges to a Poisson process with rate 1, as $n \to \infty$, if we scale time by a factor $n$, because of Lemma 1. Furthermore, (49) also tells us that for this case, if we scale time by a factor $n$, we expect to see exponential inter-transmission times with rate 1. So these results are in good agreement with each other.

### 6.2 Discussion of results

We briefly discuss the implications of our results. First, Equation (36) gives us insight into the effects of the short-listen problem. We found that without a listen-only period, the number of transmissions per time unit scales as $\mathcal{O}(\sqrt{n})$, due to the short-listen problem. However, by introducing a listen-only period, the expected number of transmissions per interval is bounded by $k/\eta$: see (37). Hence, having $\eta = \frac{1}{2}$ as suggested in [13] reduces the number of redundant transmissions and resolves the short-listen problem.

However, one could also consider using lower values of $\eta$. For example, one could consider setting $\eta = \frac{1}{4}$. We know this roughly doubles the expected number of redundant transmissions, but it could also double the propagation speed. Indeed, whenever a node gets updated, it only has to wait for a period of at least $\tau_l/4$ before propagating the update, as opposed to a period of $\tau_l/2$. Hence, the choice of $\eta$ involves a trade-off. Lowering $\eta$ increases not only the propagation speed, but also the number of transmissions. In order to get a full understanding of this trade-off, more knowledge is needed on the propagation speed of the algorithm. Combined with results from this paper, this knowledge could provide guidelines for how to optimally set the Trickle parameters and the length of the listen-only period.

Second, Eqs. (12), (31), (42), (43), (46), and (49) provide insight in the distributions of inter-transmission times. In our derivations, we have assumed that transmissions are instantaneous. However, in reality, transmissions take some time, although this time is generally short compared to the interval length $\tau_h$. Using the inter-transmission time distributions, one can estimate the probability that transmissions would overlap in time and hence interfere. Potentially, one can use this knowledge to optimize the medium access protocol and the capacity of the wireless network. Therefore, the impact of having non-instantaneous transmissions on the performance of Trickle is an interesting topic for future research.

### 7 Simulation results

The model we have developed assumes that the process of nodes attempting to broadcast is Poisson with rate $n$. Lemma 1 shows that this is true as $n$ grows large. We now investigate how well this assumption holds in networks with only a few nodes. In order to do so, we compare our analytic results for the message count and the distribution of inter-transmissions times with simulations done in Mathematica. We simulate a lossless single-cell network using the Trickle algorithm, where transmissions occur instantaneously.

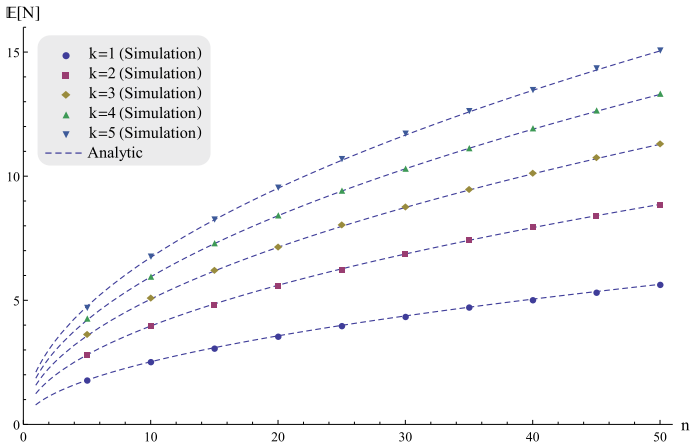

**Fig. 3** Mean number of transmissions per interval for $\eta = 0$

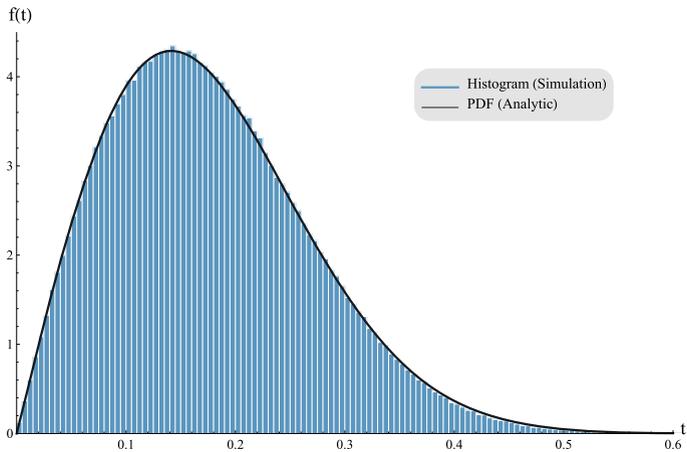

**Fig. 4** Density plot of inter-transmission times for $\eta = 0$, $k = 1$, and $n = 50$

In Fig. 3, we compare simulation results for the case $\eta = 0$ with the analytic results from Eq. (36). For each combination of $k$ and $n$, we simulate a network consisting of $n$ nodes using the Trickle algorithm and average over $10^3$ runs of 100 virtual time units. Each run, we use a different interval skew chosen uniformly at random for each of the nodes. We see that the simulation results and the analytic results coincide very well, even for small cell-sizes, and that the analytic results provide a conservative estimate for the mean number of transmissions per interval.

In Fig. 4, we compare simulation results for the case $\eta = 0$, $k = 1$, and $n = 50$ with the analytic results from Eq. (12). We show a histogram of the observed inter-transmission times obtained from $10^3$ runs of 100 virtual time units and the probability density function from (12). We see a very good match between the analytic result and simulations, even though we consider a network consisting of only 50 nodes.

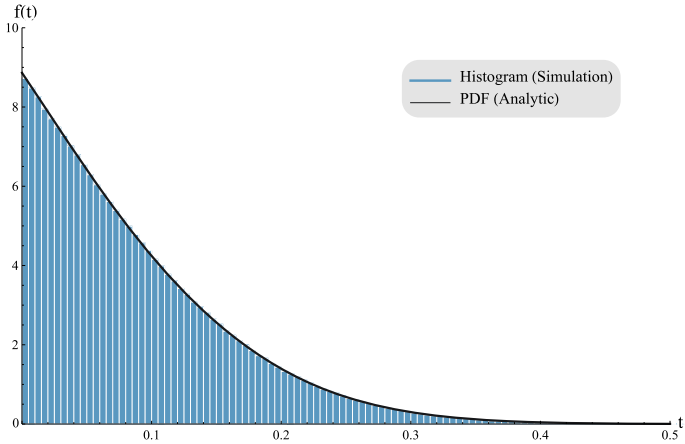

**Fig. 5** Density plot of inter-transmission times for $\eta = 0$, $k = 3$, and $n = 50$

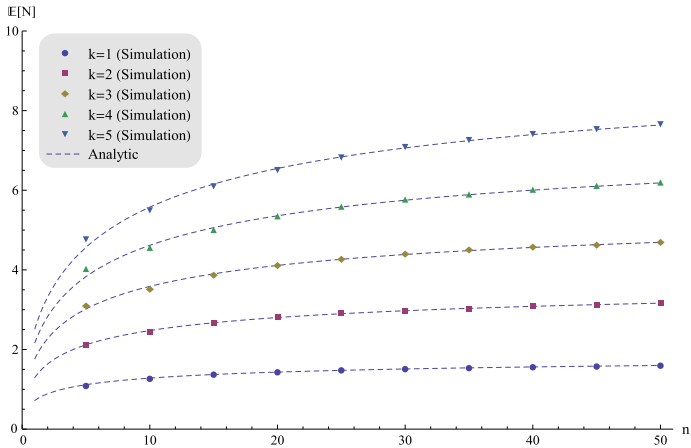

**Fig. 6** Mean number of transmissions per interval for $\eta = \frac{1}{2}$

In Fig. 5, we compare (32) with simulations for the case $\eta = 0$, $k = 3$, and $n = 50$. Again, both results are in good agreement with each other, despite the relatively small size of the network. Note, also, that the asymptotically exponential behavior from (49) can already be seen in the density plot.

In Fig. 6, we compare simulation results for the case $\eta = \frac{1}{2}$ with analytic results from Eq. (37). Here, also, we find that the analysis gives an accurate estimate for the mean number of transmissions in small networks. From our analysis, we also know that $\mathbb{E}[N^{(k,n)}]$ should converge to $2k$ as $n$ grows large. From the graph, we see that this convergence is quite slow.

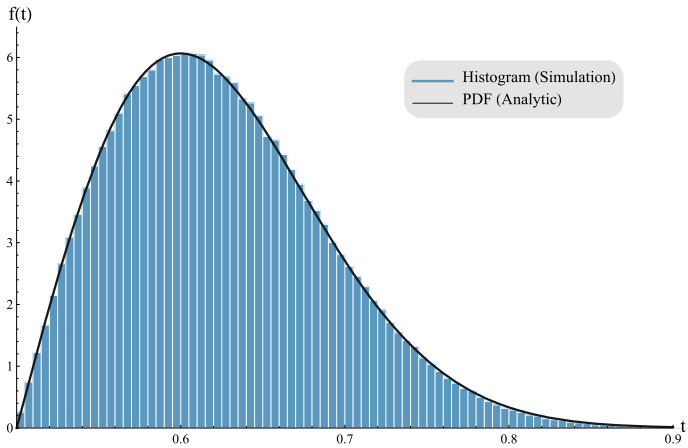

**Fig. 7** Density plot of inter-transmission times for $\eta = \frac{1}{2}$, $k = 1$, and $n = 50$

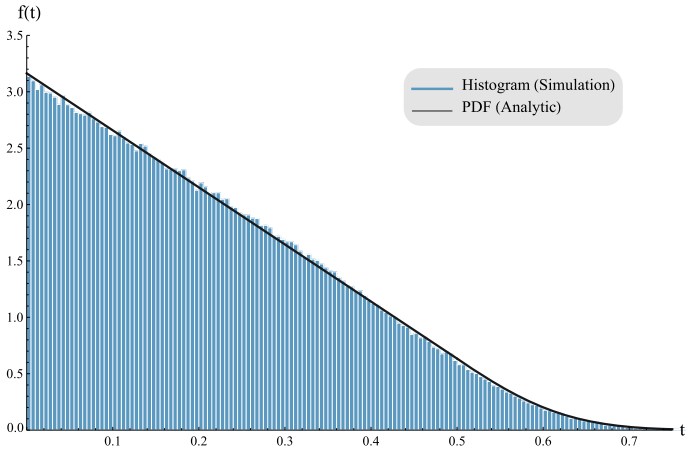

**Fig. 8** Density plot of inter-transmission times for $\eta = \frac{1}{2}$, $k = 3$, and $n = 50$

In Figs. 7 and 8, we compare simulation results for $k = 1$ and $k = 3$, respectively, with the analytic results from Eqs. (12) and (32), where $\eta = \frac{1}{2}$ and $n = 50$. Like before, we see a very good match between the analytic results and simulations. Note, that the asymptotic behavior from (41) can already be recognized in the density plot of Fig. 8.

Additionally, we have simulated more realistic network settings in the OMNeT++ simulation tool. As mentioned in Sect. 6.2, in reality, transmissions do not occur instantaneously but take some time $M$; hence, collisions can occur. To investigate the effect of not having instantaneous and possibly colliding transmissions, we have simulated the same scenarios as we did with Mathematica and compared the results. The simulator incorporates IEEE standard 802.15.4.

We find that the OMNeT++ and Mathematica simulations produce nearly indistinguishable results, which is not surprising. Since $\tau_h$ tends to be very large compared to a transmission length $M$, the probability of nodes trying to broadcast simultaneously tends to be very small. Furthermore, since we are dealing with a single-cell network, the CSMA/CA protocol prevents all collisions in the OMNeT++ simulations. Therefore, both simulators provide nearly the same results. Since the Mathematica simulations are computationally less demanding and much more easily reproducible, we have omitted the OMNeT++ simulation data.

## 8 Multi-cell network

Suppose now that we have a network consisting of $n^2$ nodes placed on a square grid, where not all nodes are able to directly communicate with each other. Instead, each node has a fixed transmission range $R$, which means that when a node sends a message, only nodes within a distance $R$ of the broadcaster receive the message. While a fully-fledged analysis of multi-cell networks is beyond the scope of the present paper, we now briefly examine how the results for the single-cell scenario obtained in Sect. 6 can be leveraged to derive a useful approximation. We denote by $N_{MC}^{(k,n,R)}$ the number of transmissions during an interval of length $\tau_h$ for a given threshold value $k$ in a cell consisting of $n \times n$ nodes with broadcasting range $R$. Hence, we are interested in determining the behavior of $\mathbb{E}\left[N_{MC}^{(k,n,R)}\right]$.

### 8.1 Approximation

We will use the analytical results from Sect. 6 to develop an approximation for the expected number of transmissions per interval in multi-cell networks. Let us denote by $S(R)$ the number of nodes a single broadcasting node having a broadcasting range of $R$ can reach, i.e., the size of a single broadcasting-cell. Heuristically, we can reason as follows. We have an $n \times n$ grid consisting of approximately $n^2/S(R)$ distinct non-overlapping broadcasting-cells. Assuming each cell behaves independently of the others, we can approximate the expected number of transmissions per interval in the multi-cell case as follows:

$$\mathbb{E}\left[N_{MC}^{(k,n,R)}\right] \approx \frac{n^2}{S(R)} \mathbb{E}\left[N^{(k,S(R))}\right] \approx \frac{n^2}{S(R)} \frac{C_{(k+1,S(R))}}{C_{(k,S(R))}}. \tag{50}$$

Using the fact that $S(R) \sim \pi R^2$ and Eqs. (36) and (37), we can get more insight into the behavior of our approximation. For $\eta = 0$, we have that as $R$ grows large:

$$\frac{n^2}{S(R)} \mathbb{E}\left[N^{(k,S(R))}\right] \approx \sqrt{\frac{2}{\pi}} \frac{n^2}{R} \frac{\Gamma\left[\frac{k+1}{2}\right]}{\Gamma\left[\frac{k}{2}\right]}.$$

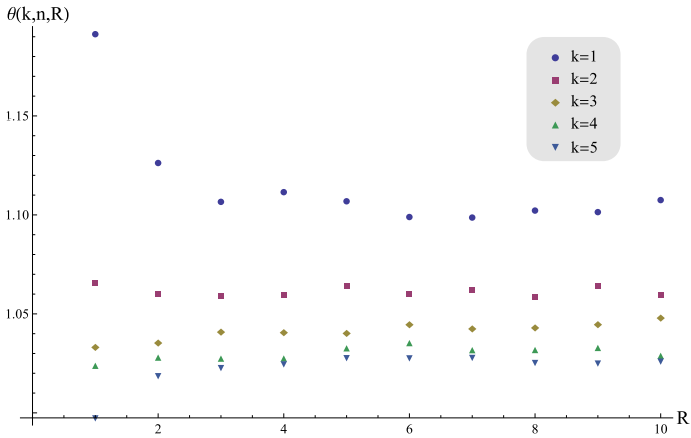

**Fig. 9** Ratio of simulated transmission count and approximation for $\eta = 0$

Similarly, when a listen-only period is used, i.e., $\eta > 0$, we have that as $R$ grows large:

$$\frac{n^2}{S(R)} \mathbb{E}\left[N^{(k,S(R))}\right] \approx \frac{n^2}{R^2} \frac{k}{\pi \eta}.$$

Consequently, we see how the short-listen problem potentially plays a role in multi-cell networks.

### 8.2 Simulations

We compare the approximation with simulations in order to evaluate its accuracy. We do so by plotting the ratio

$$\theta(k, n, R) = \mathbb{E}\left[N_{MC}^{(k,n,R)}\right] / \left(\frac{n^2}{S(R)} \frac{C_{(k+1,S(R))}}{C_{(k,S(R))}}\right). \tag{51}$$

In Mathematica, we simulate a network consisting of $50 \times 50$ nodes placed on a grid for several values of $k$ and $R$. For each combination of $k$ and $R$, we simulate a lossless multi-cell network for 100 virtual time units. Each run, we use a different randomly chosen interval skew for the nodes. We use toroidal distance in order to cope with edge effects.

In Fig. 9, we show a plot of (51) for the case $\eta = 0$. We see that the estimate given in Eq. (50) tends to slightly underestimate the number of transmissions per interval. However, the approximation is accurate within a factor 1.2. Furthermore, as $k$ increases, the approximation becomes more accurate. We also note that for fixed $k$, the accuracy remains fairly constant as $R$ grows.

Finally, let us consider the effect of a listen-only period. In Fig. 10, we show a plot of (51) for the case $\eta = \frac{1}{2}$. We see again that the estimate given in Eq. (50) tends to underestimate the number of transmissions per interval, more than for the case $\eta = 0$. The error slowly increases with the transmission range $R$. However, as $k$ increases, the approximation becomes more accurate.

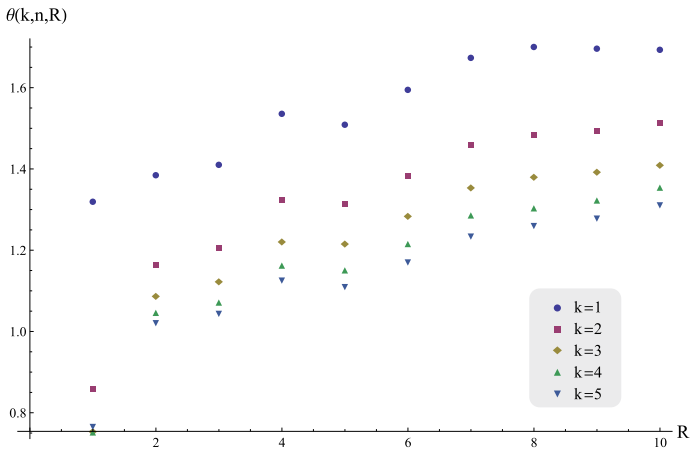

**Fig. 10** Ratio of simulated transmission count and approximation for $\eta = \frac{1}{2}$

*Remark 3* In [11], a method is presented for estimating the message count in synchronized networks. It is shown that this method also gives an accurate approximation for unsynchronized networks with $\eta = \frac{1}{2}$. However, the method is not suitable for smaller values of $\eta$ and assumes a uniformly random spatial distribution of the nodes. Hence, if one is interested in listen-only periods of different length or regular network topologies, our approximation is preferable.

*Remark 4* Given the way we derived our approximation, one can conclude that its performance strongly depends on the fact that we are considering a network which is very homogeneous in terms of node density. For networks that are more heterogeneous in terms of node density, one cannot easily decide what value to take for $S(R)$; hence, approximating the message count becomes difficult. Moreover, as shown in [22], Trickle's performance strongly depends on the network topology, which makes approximating the message count an even more difficult task.

## 9 Conclusion

In this paper, we presented a generalized version of the Trickle algorithm with a new parameter $\eta$, which allows us to set the length of a listen-only period. This parameter can greatly increase the speed at which the algorithm can propagate updates, while still controlling the number of transmissions. Furthermore, we have shown that this parameter influences how the transmission load is distributed among nodes in a network. We then presented a mathematical model describing how the message count and inter-transmission times of the Trickle algorithm depend on its various parameters.

We showed that the broadcasting process in a single-cell network can be modeled as a Markov chain and that this chain falls under a special class of Markov chains, closely related to residual lifetimes. This class of Markov chains was then analyzed and its stationary distribution derived. For our model, these results lead to the distribution function of inter-transmission times and the joint distribution of consecutive inter-transmission times. We showed how they depend on the redundancy constant $k$, the network-size $n$, and the length of a listen-only period $\eta$. We also investigated their asymptotic behavior as $n$ and $k$ go to infinity. These distributions give insight in the energy-efficiency of Trickle and the probability that nodes try to broadcast simultaneously. These insights contribute to optimizing the design and usage of the Trickle algorithm.

Specifically, we showed that in a network without a listen-only period, the expected number of transmissions grows as $\mathcal{O}(\sqrt{n})$, proving a conjecture from [13], and we identified that the pre-factor is $\sqrt{2}\Gamma\left[\frac{k+1}{2}\right]/\Gamma\left[\frac{k}{2}\right]$. Additionally, we showed that, when a listen-only period is used, the number of transmissions per interval is bounded from above by $\frac{k}{\eta}$, proving a second conjecture from [13] on the scalability of Trickle.

We have also performed a simulation study in Mathematica and the OMNeT++ simulation tool. We compared our analytic results, which hold for very large networks with instantaneous transmissions, with simulation results of small and more realistic wireless networks. We found a very good match between the analytic results and the simulations.

Additionally, we used the results from the single-cell analysis to get an approximation for the message count in multi-cell networks. These results were also compared to simulation results from Mathematica and the OMNeT++ simulation tool. The approximation proved to be fairly accurate, in particular, for small values of $\eta$. A more comprehensive investigation of multi-cell networks and the influence of network topology on Trickle's performance would be interesting topics for further research.

Finally, we note that the speed at which the Trickle algorithm can disseminate updates is discussed in [15]. Combined with results from this paper, this could provide insight in how to optimally set the Trickle parameters and the length of the listen-only period. Additionally, the impact of non-instantaneous broadcasts, interference, and combining Trickle with CSMA/CA on the performance of the Trickle algorithm is discussed in [20].

**Acknowledgments** The authors would like to thank Guido Janssen for useful discussions regarding the integral equations presented in this paper. Additionally, they are grateful to Marc Aoun for his contributions to the performed OMNeT++ simulations. The research of the third author was done in the framework of the IAP BESTCOM program, funded by the Belgian government.

## Appendix: Proof of Lemma 1

In order to prove Lemma 1, we will use the following theorem (see [5], Proposition 11.2.VI).

**Theorem 3** *Let M be a simple stationary point process on $X = \mathbb{R}$ with finite intensity $\lambda$, and let $M_n$ denote the point process obtained by superposing n independent replicates of M and dilating the scale of X by a factor n. Then as $n \to \infty$, $M_n$ converges weakly to a Poisson process with parameter measure $\lambda \mu_L(\cdot)$, where $\mu_L(\cdot)$ denotes the Lebesgue measure on $\mathbb{R}$.*

Let $N$ be the process of an arbitrary node's broadcasting moments. Then evidently the superposition of $n$ of these processes $N_n$ gives us the process of broadcasting attempts by all the nodes in our cell. Hence, if we show that $N$ is a simple stationary point process with finite intensity $\lambda = 1$, we can apply Theorem 3, resulting in Lemma 1.

Let $\{B_i\}$, with $i \in \mathbb{Z}$, be the sequence of broadcasting times for an arbitrary node defining our process $N$. Then if we denote by $S$ the interval skew and assume it is uniform, i.e., $S \sim U[0, 1]$, we can write

$$B_i \sim (S + i) + T_i, i \in \mathbb{Z}.$$

Here $T_i$ represents the timer of the corresponding node in the $i$th interval; thus, $T_i \sim U[\eta, 1]$. It is easy to see that $N$ is a simple point process with intensity $\lambda = 1$.

It remains to show stationarity. Let $N^s$ be the point process resulting from applying a shift operator to the process $N$. That is, if $\{B_i^s\}$, with $i \in \mathbb{Z}$, is the sequence of events for the shifted process $N^s$, we have

$$B_i^s \sim (S + i) + T_i + s, i \in \mathbb{Z}.$$

We show that $\{B_i\}$ and $\{B_i^s\}$, with $i \in \mathbb{Z}$, have the same distribution, which means that the processes $N$ and $N^s$ coincide in distribution, implying stationarity. We first write

$$\begin{aligned}\{B_i^s\}_{i \in \mathbb{Z}} &= \{(S + i) + T_i + s\}_{i \in \mathbb{Z}} \\ &= \{(S + s - \lfloor S + s \rfloor) + (i + \lfloor S + s \rfloor + T_i)\}_{i \in \mathbb{Z}}\end{aligned}$$

Now note that $(S + s - \lfloor S + s \rfloor) \sim U[0, 1] \sim S$. Furthermore, since the $T_i$ are independent and identically distributed, $(i+\lfloor S+s \rfloor+T_i) \sim \left(i + \lfloor S + s \rfloor + T_{i+\lfloor S+s \rfloor}\right)$. Moreover, $\left\{i + \lfloor S + s \rfloor + T_{i+\lfloor S+s \rfloor}\right\}_{i \in \mathbb{Z}} \sim \{i + T_i\}_{i \in \mathbb{Z}}$. This allows us to conclude

$$\{B_i^s\}_{i \in \mathbb{Z}} \sim \{B_i\}_{i \in \mathbb{Z}}.$$

Consequently, Theorem 3 applies and Lemma 1 follows.